\documentclass[%
 reprint,
showpacs,preprintnumbers,
 amsmath,amssymb,
 aps,
]{revtex4-1}

\usepackage{graphicx}
\usepackage{dcolumn}
\usepackage{bm}

\begin{document}


\title{Detection of monopoles and vortices in SU($2$) Yang-Mills theory}

\author{N. Karimimanesh}
 \altaffiliation{karimi.n@ut.ac.ir}
\author{S. Deldar}%
 \email{sdeldar@ut.ac.ir}
\affiliation{Department of Physics, University of Tehran,\\
P.O. Box 14395-547, Tehran 1439955961, Iran}



\begin{abstract}
Motivated by the correlated monopoles and vortices observed in lattice QCD, using Cho decomposition method and two successive gauge transformations which lead to the observation of configurations that include both monopoles and center vortices simultaneously, we obtain a Lagrangian density that explicitly indicates these two magnetic defects and the interaction between them. 

\begin{description}
\item[PACS numbers] 14.80.Hv, 12.38.Aw, 12.38.Lg, 12.39.Pn
\end{description}
\end{abstract}

\pacs{Valid PACS appear here}
\maketitle


\section{\label{sec:level}Introduction}

One of the unsolved and challenging problems in particle physics is the confinement of quarks in Quantum Chromodynamics. There exist various suggestions to approach this difficult problem. There are many articles which discuss about the magnetic objects that play significant roles in confinement of quarks. Some of them are listed in references \cite{gbook} to \cite{e}. Monopoles and vortices are among the major candidates which have been studied to describe confinement of color sources.   
 
In the absence of matter fields, two main methods have been presented to extract magnetic monopole degrees of freedom in the YM (Yang-Mills) theories. The first one is the Abelian projection proposed by 't Hooft \cite{hooft} and the second one is a field decomposition method which is introduced by Duan, Ge, Cho, Faddeev and Niemi \cite{duan, cho1, cho2, cho3, fadev, shaban}.

On the other hand, some lattice calculations show that line-like (surface-like) objects in three (four) dimensions \cite{vor} are responsible for the phenomenon of confinement. A method called Center Projection has been proposed \cite{faber} to examine the role of center vortices in the Yang-Mills lattice gauge configurations, which has been very successful.
However, scenarios which are solely based on either monopoles or center vortices are not able to describe all the expected details of the confining potential between color sources. The confining potential should be linear as well as proportional to the Casimir scaling at intermediate distances. In addition, N-ality dependence should be observed for all representations at large distances.  The advantage of the center vortex picture versus the monopole picture is that it explains N-ality and Casimir scaling dependence. On the other hand, the monopole picture along with Abelian dominance exhibits some numerical success. (for a review, see ref. \cite{greens}). A clever idea to overcome these problems and to reconcile these two candidates of confinement is to present a theory which establishes a correlation between center vortices and monopoles. The corresponding configurations have been supported by lattice simulations \cite{greensite, pepe, zakharov} where center vortices have been observed to end at monopole world-lines.

Some phenomenological attempts have been done in using both vortices and monopoles to describe the color confinement.
For example, in reference \cite{oxman}, by two successive gauge transformations and by means of a careful treatment of Cho decomposition, some configurations which simultaneously include monopoles and vortices have been discussed. A nontrivial transformation leads to the appearance of monopoles, and an SU($2$) gauge transformation that is not single valued along a closed loop was shown to be responsible for the existence of center vortices. In another paper \cite{val}, a Yang-Mills-Higgs model containing three external adjoint Higgs fields was presented. This model leads to Z(2) vortices and the junctions which are formed by a pair of vortices, are attached to a monopole-like configurations. The ultimate goal of the model was to use it to describe hybrid hadrons. Some other works \cite{shabani} and \cite{4Densem} in reconciling vortices and monopoles have been reported. 

In this article, motivated by the method presented in reference \cite{oxman} and by comparison with Abelian Projection scenario which has been discussed in reference \cite{ichie}, we obtain a Lagrangian density for QCD vacuum in the confining regime and discuss explicitly the contributions of various defects such as vortex, anti vortex, monopoles, chains and 
eventually the interaction between monopoles and vortices. 
In Section \ref{sec:level1}, by limiting ourselves to SU($2$) Yang-Mills fields, we carefully review the Cho decomposition method. In section \ref{sec2}, magnetic monopoles are discussed by a nontrivial gauge transformation which results to a local color frame including monopoles. Then, the gauge transformed field strength tensor is calculated and its various terms
are interpreted. In section \ref{sec4}, using a gauge transformation that is not single valued along any closed loop, center vortices are extracted. In 
section \ref{sec3}, it is shown that magnetic monopoles and vortices can simultaneously appear as two defects in the color frame $\hat{n}_a, a=1,2,3$. 
Then, we obtain a Lagrangian density for correlated monopoles and vortices. Finally, in Section \ref{sec5}, we present our conclusions. In this article, we study 
SU($2$) gauge group for simplicity. With some more efforts, it can be extended to higher gauge groups like SU($3$).

\section{\label{sec:level1}Cho decomposition method}

In this section we review Cho decomposition method for the SU($2$) gauge group. For higher gauge groups like SU($3$), that have SU($2$) subgroups, field decomposition can be written in a Weyl symmetric way \cite{frank18}. Our main discussion in this paper is about SU($2$) group and we will not go into SU($3$). 
\\

The Yang-Mills action of SU($2$) is defined as the following
\begin{equation}
S_{YM}=\frac{1}{2}\int d^4x\  tr(F_{\mu\nu}F_{\mu\nu})\      \   ,    \         F_{\mu\nu}=F_{\mu\nu}^a T_a,       
\end{equation}
\begin{equation}
 F_{\mu\nu}^a=\partial_\mu A_\nu^a -\partial_\nu A_\mu^a+g\epsilon^{abc} A_\mu^b\times A_\nu^c,
\end{equation}
where $g$ is the coupling constant, $T^a=\frac{\tau^a}{2}, a=1,2,3$ are generators of SU($2$), $\tau^a$ are the Pauli matrices, and $\epsilon^{abc}$ is the Levi-Civita symbol. The field strength tensor can be written in terms of the gauge fields $\vec{A}_\nu$
\begin{equation}
\vec{F}_{\mu\nu}=\partial_\mu\vec{A}_\nu-\partial_\nu\vec{A}_\mu+g\vec{A}_\mu\times\vec{A}_\nu, \label{f}
\end{equation}
\begin{equation}
 \vec{A}_\mu =A_\mu^a \hat{e}_a\   \    ,    \      \vec{F}_{\mu\nu}=F^a_{\mu\nu}\hat{e}_a,
\end{equation}
where $\hat{e}_a$ is the basis in the color space.\\

A general local frame is considered in the color space, $\hat{n}_a , a=1,2,3$ which can be defined by means of an orthogonal local transformation $R$
\begin{equation}
\hat{n}_a=R\hat{e}_a\   \   ,   \    R\in SO(3).
\end{equation}
This frame is used to represent the gauge field $\vec{A}_\mu$ in terms of Cho decomposition elements
\begin{equation}
\vec{A}_\mu=A_\mu^{(n)}\hat{n}-\frac{1}{g}\hat{n}\times\partial_\mu \hat{n}+\vec{X}_\mu^{(n)}\   \   ,   \   \hat{n}.\vec{X}_\mu^{(n)}=0,\label{Cho}
\end{equation}
\begin{equation}
\hat{n}_a.\hat{n}_b=\delta_{ab}\   \   ,   \   a,b=1,2,3\   \   ,   \   \hat{n}\equiv\hat{n}_3,
\end{equation}
where $A_\mu^{(n)}$ is called the electric potential or the Abelian component of $\vec{A}_\mu$ directed along $\hat{n}\equiv\hat{n}_3$ and $\vec{X}_\mu^{(n)}$ is a gauge covariant potential orthogonal to $\hat{n}$ called the valence potential.\\

The restricted potential is defined as, (see \cite{cho2000} and its references)
\begin{equation}
\hat{A}_\mu =A_\mu^{(n)}\hat{n}-\frac{1}{g}\hat{n}\times\partial_\mu\hat{n}.\label{restrict}
\end{equation}
Using $\hat{A}_\mu$ of Eqn. (\ref{restrict}), one can easily calculate the field strength tensor
\begin{equation}
\vec{F}_{\mu\nu}=(F_{\mu\nu}^{(n)}+H_{\mu\nu}^{(n)})\hat{n},\label{f-res}
\end{equation}
where
\begin{equation}
F_{\mu\nu}^{(n)}=\partial_\mu A_\nu^{(n)}-\partial_\nu A_\mu^{(n)},
\end{equation}
\begin{equation}
H_{\mu\nu}^{(n)}=-\frac{1}{g}\hat{n}.(\partial_\mu\hat{n}\times\partial_\nu\hat{n}).\label{hmunu}
\end{equation}
Eqn. (\ref{f-res}) shows that $\vec{F}_{\mu\nu}$ is parallel to $\hat{n}$, and it is made of two parts: $F_{\mu\nu}^{(n)}$ which is called the electric strength tensor and $H_{\mu\nu}^{(n)}$ which is called the magnetic strength tensor.

Choosing a hedgehog configuration for $\hat{n}$ and choosing $A_\mu^{(n)}=0$, one can obtain a Wu-Yang monopole \cite{wu}
\begin{equation}
\hat{n}=\hat{r}=\begin{pmatrix}
\sin\theta\cos\varphi\\
\sin\theta\sin\varphi\\
\cos\theta
\end{pmatrix},\label{hedg}
\end{equation}
Using Eqn. (\ref{hedg}) in Eqn. (\ref{hmunu}), $H_{\mu\nu}^{(n)}$ can be written in the following format
\begin{equation}
H_{\mu\nu}^{(n)}=\partial_\mu C_\nu^{(n)}-\partial_\nu C_\mu^{(n)},
\end{equation}
where
\begin{equation}
C_\mu^{(n)}=\frac{1}{g}\cos\theta\partial_\mu\varphi,\label{mag}
\end{equation}
$C_\mu^{(n)}$ is called magnetic potential.

Therefore, using Cho decomposition method and the restricted potential of Eqn. (\ref{restrict}), the Wu-Yang monopole is extracted.

In order to show that center vortices can be obtained by Cho decomposition method, we do not limit ourselves to the restricted section of the field, and the role of each of the three fields in Eqn. (\ref{Cho}) is examined. 
Using $\vec{A}_\mu$ of Eqn. (\ref{Cho}) obtained from Cho decomposition method to compute the field strength tensor, we get
\begin{equation}
\vec{F}_{\mu\nu}=(F_{\mu\nu}^{(n)}+H_{\mu\nu}^{(n)}+K_{\mu\nu})\hat{n}+\vec{G}_{\mu\nu}+\vec{L}_{\mu\nu},\label{F}
\end{equation}
\begin{equation}
F_{\mu\nu}^{(n)}=\partial_\mu A_\nu^{(n)}-\partial_\nu A_\mu^{(n)}\   \   ,   \   H_{\mu\nu}^{(n)}=-\frac{1}{g}\hat{n}.(\partial_\mu\hat{n}\times\partial_\nu\hat{n}),\label{hh}
\end{equation}
\begin{equation}
K_{\mu\nu}=g\hat{n}.(\vec{X}_\mu\times\vec{X}_\nu)\   \   ,   \   \vec{X}_\mu =X_\mu ^1\hat{n}_1+X_\mu ^2\hat{n}_2,\label{Xmu}
\end{equation}
\begin{equation}
\begin{split}
\vec{G}_{\mu\nu}&=\hat{D}_\mu\vec{X}_\nu^{(n)}-\hat{D}_\nu\vec{X}_\mu^{(n)},\\ \hat{D}_\mu\vec{X}_\nu^{(n)}&=\partial_\mu\vec{X}_\nu^{(n)}+g\hat{A}_\mu\times\vec{X}_\nu^{(n)},\label{dx}
\end{split}
\end{equation}
\begin{equation}
\vec{L}_{\mu\nu}=-\frac{1}{g}\hat{n}\times[\partial_\mu ,\partial_\nu]\hat{n}=L_{\mu\nu}^1\hat{n}_1+L_{\mu\nu}^2\hat{n}_2.\label{lmu}
\end{equation}

It is clear that $\vec{L}_{\mu\nu}$ which is orthogonal to $\hat{n}$, is concentrated on two-dimensional surfaces. It could be nontrivial only for local frames containing defects in the color direction $\hat{n}$. The tensor $\vec{G}_{\mu\nu}$ have been computed in \cite{cho1, cho2, cho3}, \cite{cho2000, lee, bae, pak}: $\vec{G}_{\mu\nu}=G_{\mu\nu}^1\hat{n}_1+G_{\mu\nu}^2\hat{n}_2$, being orthogonal to $\hat{n}$.

In the presence of singularities, this result may remain unchanged. For simplicity, a new “Abelianized” form for the field strength tensor is defined: $\partial_\mu C_\nu^{(n)}-\partial_\nu C_\mu^{(n)}$ which should be revised when the gauge fields containing defects.

Using $\hat{A}_\mu$ of Eqn. (\ref{restrict}) and  $\vec{X}_\mu$ of Eqn. (\ref{Xmu}) in Eqn. (\ref{dx}), 
and defining
\begin{equation}
C_\mu^{(n)}=-\frac{1}{g}\hat{n}_1.\partial_\mu\hat{n}_2,\label{cmu}
\end{equation}
the covariant derivative of $\vec{X}_\mu^{(n)}$ is obtained as the following
\begin{equation}
\begin{split}
\hat{D}_\mu\vec{X}_\nu^{(n)}&=[\partial_\mu X_\nu ^1-g(A_\mu^{(n)}+C_\mu^{(n)})X_\nu ^2]\hat{n}_1\\
&+[\partial_\mu X_\nu ^2-g(A_\mu^{(n)}+C_\mu^{(n)})X_\nu ^1]\hat{n}_2.
\end{split}
\end{equation}
Now, we can redefine $H_{\mu\nu}^{(n)}$ in terms of $C_\mu^{(n)}$
\begin{equation}
H_{\mu\nu}^{(n)}=\partial_\mu C_\nu^{(n)}-\partial_\nu C_\mu^{(n)}+D_{\mu\nu}^{(n)},\label{hmu}
\end{equation}
where $D_{\mu\nu}^{(n)}$ is obtained to be
\begin{equation}
D_{\mu\nu}^{(n)}=\frac{1}{g}\hat{n}_1.[\partial_\mu ,\partial_\nu]\hat{n}_2.\label{hc}
\end{equation}
In the following we show how $D_{\mu\nu}^{(n)}$ is obtained. From the definition of Eqn. (\ref{cmu}),
\begin{equation}
\begin{split}
\partial_\mu C_\nu^{(n)}&-\partial_\nu C_\mu^{(n)}=-\frac{1}{g}[\partial_\mu(\hat{n}_1.\partial_\nu\hat{n}_2)-\partial_\nu(\hat{n}_1.\partial_\mu\hat{n}_2)]\\
&=-\frac{1}{g}(\partial_\mu\hat{n}_1.\partial_\nu\hat{n}_2-\partial_\nu\hat{n}_1.\partial_\mu\hat{n}_2)-\frac{1}{g}\hat{n}_1.[\partial_\mu ,\partial_\nu]\hat{n}_2,\label{cc}
\end{split}
\end{equation}
\begin{equation}
\begin{split}
\partial_\mu\hat{n}_1.\hat{n}_1=0&\Rightarrow\partial_\mu\hat{n}_1=\alpha_\mu^1\hat{n}_2+\beta_\mu^1\hat{n},\\
\partial_\mu\hat{n}_2.\hat{n}_2=0&\Rightarrow\partial_\mu\hat{n}_2=\alpha_\mu^2\hat{n}_1+\beta_\mu^2\hat{n},\\
\end{split}
\end{equation}
\begin{equation}
\begin{split}\hat{n}=\hat{n}_1\times\hat{n}_2\Rightarrow\partial_\mu\hat{n}&=\partial_\mu\hat{n}_1\times\hat{n}_2+\hat{n}_1\times\partial_\nu\hat{n}_2\\
&=-\beta_\mu^1\hat{n}_1-\beta_\mu^2\hat{n}_2.\label{n*n}
\end{split}
\end{equation}
By replacing Eqn. (\ref{n*n}) in Eqn. (\ref{cc}) 
\begin{equation}
\partial_\mu C_\nu^{(n)}-\partial_\nu C_\mu^{(n)}=-\frac{1}{g}(\beta_\mu^1\beta_\nu^2-\beta_\nu^1\beta_\mu^2)-\frac{1}{g}\hat{n}_1.[\partial_\mu ,\partial_\nu]\hat{n}_2.\label{b}
\end{equation}
On the other hand, by replacing Eqn. (\ref{n*n}) in Eqn. (\ref{hh})
\begin{equation}
\begin{split}
\partial_\mu\hat{n}\times\partial_\nu\hat{n}&=(-\beta_\mu^1\hat{n}_1-\beta_\mu^2\hat{n}_2)\times(-\beta_\nu^1\hat{n}_1-\beta_\nu^2\hat{n}_2)\\
&=(\beta_\mu^1\beta_\nu^2-\beta_\nu^1\beta_\mu^2)\hat{n}
\end{split}
\end{equation}
\begin{equation}
\hat{n}.(\partial_\mu\hat{n}\times\partial_\nu\hat{n})=(\beta_\mu^1\beta_\nu^2-\beta_\nu^1\beta_\mu^2).\label{aa}
\end{equation}
By comparing Eqn. (\ref{b}) and Eqn. (\ref{aa}), $D_{\mu\nu}^{(n)}$ in Eqn. (\ref{hc}) is obtained.

For a framework that has no singularity in two-dimensional surface orthogonal to $\hat{n}$, $D_{\mu\nu}^{(n)}$ is zero. In fact, it can be stated that if $C_\mu^{(n)}$ is the magnetic monopole potential in the third color direction, then the associated Dirac worldsheet is localized on its orthogonal surface. Therefore, when $D_{\mu\nu}^{(n)}$ is non-zero, the contribution of the Dirac worldsheets should be considered .

If the frames were regular, we would have $\vec{L}_{\mu\nu}=\vec{0}$, $D_{\mu\nu}^{(n)}=0$, and substituting them in Eqn. (\ref{F}), we would obtain the Abelianized form given in \cite{cho2000, lee, bae, pak}
\begin{equation}
\begin{split}
\vec{F}_{\mu\nu}&=(\partial_\mu(A_\nu^{(n)}+C_\nu^{(n)})-\partial_\nu(A_\mu^{(n)}+C_\mu^{(n)})+K_{\mu\nu})\hat{n}\\
&+\hat{D}_\mu \vec{X}_\nu^{(n)}-\hat{D}_\nu \vec{X}_\mu^{(n)}.\label{fst}
\end{split}
\end{equation}
Therefore, taking into account a general configuration containing monopoles, the associated Dirac worldsheets, and center vortices and introducing a local frame containing these defects, the field strength tensor is calculated by Cho decomposition method. In the next sections, using the results of this section, possible defects and the associated consequences are studied. 

\section{Monopoles in SU($2$) gauge group}\label{sec2}

In the previous section, it is shown that magnetic monopoles can be extracted using Cho decomposition method and by attributing the defects to the local color frame. In order to make contact with Cho decomposition, using a nontrivial gauge transformation \cite{oxman}, a local color frame is defined and the monopoles are studied as defects of this local color frame. At the end, we interpret the components of the ultimate field strength tensors.

In general, the Yang-Mills theory can be affected in two different ways from gauge transformations. Under regular gauge transformations $S\in SU(N)$, the Yang-Mills action is invariant and
\begin{equation}
\begin{split}
\vec{A}_\mu ^S.\vec{T}&=S\vec{A}_\mu.\vec{T}S^{-1}+\frac{i}{g}S\partial_\mu S^{-1},\\
\vec{F}_{\mu\nu}^S.\vec{T}&=S\vec{F}_{\mu\nu}.\vec{T}S^{-1}.
\end{split}
\end{equation}
It should be noted that by a regular gauge transformation, the transformation and its derivative lack any kind of singularity.
\\
Under nontrivial gauge transformations, the Yang-Mills gauge field and the field strength change as the following
\begin{equation}
\vec{A}_\mu ^U.\vec{T}=U\vec{A}_\mu.\vec{T}U^{-1}+\frac{i}{g}U\partial_\mu U^{-1},\label{a}
\end{equation}
\begin{equation}
\vec{F}_{\mu\nu}^U.\vec{T}=U\vec{F}_{\mu\nu}.\vec{T}U^{-1}+\frac{i}{g}U[\partial_\mu,\partial_\nu]U^{-1}.\label{gauge}
\end{equation}
These transformations indicate some singularities for the nontrivial gauge transformations. The singularities may appear in the transformation or its derivative. 
Because of the second term in Eqn. (\ref{gauge}), the fields $\vec{A}_\mu^U$ and $\vec{A}_\mu$ are not physically equivalent. As a result, Eqn. (\ref{a}) does not show a simple gauge transformation and it may contain monopole-like defects. 

In order to relate the above discussions to Cho decomposition, we need to define a color frame $\hat{m}_a, a=1,2,3,$ which is created by the nontrivial single valued gauge transformation $U$
\begin{equation}
UT^a U^{-1}=\hat{m}_a.\vec{T}\   \   or,   \   \hat{m}_a=R(U)\hat{e}_a.\label{ma}
\end{equation}
This transformation can be expressed in terms of the Euler angles
\begin{equation}
U=e^{-i\alpha T_3}e^{-i\beta T_2}e^{-i\gamma T_3}\   \   ,   \   R(U)=e^{\alpha M_3}e^{\beta M_2}e^{\gamma M_3},\label{39}
\end{equation}
where $M_a$ are the generators of $SO(3)$ gauge group.

For a general nontrivial gauge transformation, the second term of Eqn. (\ref{a}) is obtained as the following \cite{oxman},
\begin{equation}
\frac{i}{g}U\partial_\mu U^{-1}
=-(C_\mu^{(m)}\hat{m}+\frac{1}{g}\hat{m}\times\partial_\mu\hat{m}).\vec{T}.\label{44}
\end{equation}
where $C_\mu^{(m)}$ is defined in Eqn. (\ref{cmu}).

By applying Eqn. (\ref{44}) and
\begin{equation}
\begin{split}
U\vec{A}_\mu.\vec{T}U^{-1}&=U(A_\mu^1 T^1+A_\mu^2 T^2+A_\mu^3 T^3)U^{-1}\\
&=A_\mu^1 \hat{m}_1.\vec{T}+A_\mu^2\hat{m}_2.\vec{T}+A_\mu^3 \hat{m}.\vec{T},
\end{split}
\end{equation}
to Eqn. (\ref{a}), we obtain
\begin{equation}
\vec{A}_\mu^U.\vec{T}=[(A_\mu^3-C_\mu^{(m)})\hat{m}-\frac{1}{g}\hat{m}\times\partial_\mu \hat{m}+A_\mu^1\hat{m}_1+A_\mu^2\hat{m}_2].\vec{T}.\label{46}
\end{equation}
Defining 
\begin{equation}
A_\mu^{(m)}=A_\mu^3-C_\mu^{(m)}\   \   ,   \   \vec{X}_\mu^{(m)}=A_\mu^1\hat{m}_1+A_\mu^2\hat{m}_2,
\end{equation}
we get
\begin{equation}
\vec{A}_\mu^U=A_\mu^{(m)}\hat{m}-\frac{1}{g}\hat{m}\times\partial_\mu\hat{m}+\vec{X}_\mu^{(m)}.\label{amu}
\end{equation}
Both representations (\ref{a}) and (\ref{amu}) are equivalent in describing monopoles \cite{oxman}. Monopole-like singularities of the connection $\vec{A}_\mu$ are described in terms of
a defect located in the color direction $\hat{m}$ \cite{cho2000, lee, bae, pak}. If we choose $A_\mu^1=A_\mu^2=0$, the gauge field $\vec{A}_\mu^U$ in Eqn. (\ref{amu}) changes to the restricted potential mentioned in section \ref{sec:level1}. We recall from section \ref{sec:level1} that by considering a hedgehog configuration for $\hat{m}$ and defining $A_\mu^{(m)}=0$, a Wu-Yang monopole \cite{wu} is obtained.
\\
\subsection{Interpreting the transformed gauge field and field strength tensor}

Now, we use the gauge field $\vec{A}_\mu^U$ obtained in Eqn. (\ref{46}) and calculate the field strength tensor and interpret its different terms. Then, we discuss about the field strength tensor in terms of magnetic defects. 
  
We consider a hedgehog form for $\hat{m}$ as $\hat{m}=\hat{r}$ and calculate the components of $\vec{A}_\mu^U$. The hedgehog configuration can be obtained by choosing $\alpha=\varphi, \beta=\theta, \gamma=\varphi$ for $U$ in Eqn. (\ref{39}), where $\theta , \varphi$ are the polar angles associated to $\hat{r}$.
\begin{equation}
U=
\left(
\begin{array}{cc}
\cos\frac{\theta}{2}e^{-i\varphi}&-\sin\frac{\theta}{2}\\
\sin\frac{\theta}{2}&\cos\frac{\theta}{2}e^{i\varphi}
\end{array}
\right)\label{50}
\end{equation}
Using $U$ of Eqn. (\ref{50}) in Eqn. (\ref{ma})
\begin{equation}
\begin{split}
UT^1U^{-1}=\hat{m}_1.\vec{T}\Longrightarrow\hat{m}_1&=\begin{pmatrix}
-\sin^2\frac{\theta}{2}+\cos^2\frac{\theta}{2}\cos2\varphi\\
\cos^2\frac{\theta}{2}\sin2\varphi\\
-\sin\theta\cos\varphi
\end{pmatrix}\label{m1}\\
&=\cos\varphi\ \hat{\theta}+\sin\varphi\ \hat{\varphi}
\end{split}
\end{equation}
\begin{equation}
\begin{split}
UT^2U^{-1}=\hat{m}_2.\vec{T}\Longrightarrow\hat{m}_2&=\begin{pmatrix}
-\cos^2\frac{\theta}{2}\sin2\varphi\\
\sin^2\frac{\theta}{2}+\cos^2\frac{\theta}{2}\cos2\varphi\\
\sin\theta\sin\varphi
\end{pmatrix}\label{m2}\\
&=-\sin\varphi\ \hat{\theta}+\cos\varphi\ \hat{\varphi}
\end{split}
\end{equation}
\begin{equation}
UT^3U^{-1}=\hat{m}.\vec{T}\Longrightarrow\hat{m}=\begin{pmatrix}
\sin\theta\cos\varphi\\
\sin\theta\sin\varphi\\
\cos\theta
\end{pmatrix}=\hat{r}\label{m3}
\end{equation}
$\hat{r}$, $\hat{\theta}$ and $\hat{\varphi}$ are the unit vectors of spherical coordinates.

By replacing the above equations in Eqn. (\ref{46}), the components of $\vec{A}_\mu^U$ is obtained
\begin{widetext}
\begin{equation}
\begin{split}
(A_\mu^U)^1&=(-\sin^2\frac{\theta}{2}+\cos^2\frac{\theta}{2}\cos2\varphi)A_\mu^1-\cos^2\frac{\theta}{2}\sin 2\varphi A_\mu^2+\sin\theta\cos \varphi A_\mu^3+\frac{1}{g}\sin\varphi\partial_\mu\theta-\frac{1}{g}\sin\theta \cos\varphi\partial_\mu\varphi\\
(A_\mu^U)^2&=\cos^2\frac{\theta}{2}\sin 2\varphi A_\mu^1+(\sin^2\frac{\theta}{2}+\cos^2\frac{\theta}{2}\cos2\varphi) A_\mu^2+\sin \theta\sin\varphi A_\mu^3-\frac{1}{g}\cos\varphi\partial_\mu\theta-\frac{1}{g}\sin\theta\sin\varphi\partial_\mu\varphi\\
(A_\mu^U)^3&=-\sin\theta \cos\varphi A_\mu^1+\sin\theta\sin\varphi A_\mu^2+\cos\theta A_\mu^3-\frac{1}{g}(1+\cos\theta)\partial_\mu\varphi
\end{split}\label{aut}
\end{equation}
\end{widetext}
The first three terms in all three components are regular fields. We discuss about the remaining terms and their physical interpretation.
The last two terms in $(A_\mu^{U})^1$ are proportional to $\frac{1}{r}$ and describes the contribution of monopoles. For the static monopoles, $\partial_\mu$ can be written as $\vec{\nabla}$.
\begin{equation}
\begin{split}
\frac{1}{g}\sin\varphi\vec{\nabla}\theta&=\frac{1}{g}\frac{1}{r}\sin\varphi\ \hat{\theta} \\
-\frac{1}{g}\sin\theta \cos\varphi\vec{\nabla}\varphi&=-\frac{1}{g}\frac{1}{r}\cos\varphi\ \hat{\varphi}
\end{split}
\end{equation}
Monopoles are also found in the last two terms of $(A_\mu^{U})^2$ 
\begin{equation}
\begin{split}
-\frac{1}{g}\cos\varphi\vec{\nabla}\theta&=-\frac{1}{g}\frac{1}{r}\cos\varphi\ \hat{\theta}\\
-\frac{1}{g}\sin\theta\sin\varphi\vec{\nabla}\varphi&=-\frac{1}{g}\frac{1}{r}\sin\varphi\ \hat{\varphi}
\end{split}
\end{equation}
The last term of $(A_\mu^{U})^3$ is proportional to $\frac{1}{r}$ and it diverges at $\theta=0$. Therefore it describes a monopole attached to a Dirac string.
\begin{equation}
-\frac{1}{g}(1+\cos\theta)\vec{\nabla}\varphi=-\frac{1}{g}\frac{1}{r}\frac{1+\cos\theta}{\sin\theta}\ \hat{\varphi}
\end{equation}
We can obtain the magnetic flux corresponding to the above monopole terms which penetrates the area inside the closed contour $c(r,\theta)\equiv\{(r,\theta,\varphi)|0\leq\varphi<2\pi\}$
\begin{equation}
\begin{split}
\Phi^{\text{flux}}&=\frac{1}{g}\int_0^{2\pi}r\sin\theta d\varphi\ \hat{\varphi}\cdot(\sin\varphi\vec{\nabla}\theta-\sin\theta \cos\varphi\vec{\nabla}\varphi)T^1\\
&=\frac{1}{g}\int_0^{2\pi}-\sin\theta \cos\varphi\  d\varphi\  T^1=0
\end{split}
\end{equation}
\begin{equation}
\begin{split}
\Phi^{\text{flux}}&=-\frac{1}{g}\int_0^{2\pi}r\sin\theta d\varphi\ \hat{\varphi}\cdot(\cos\varphi\vec{\nabla}\theta+\sin\theta\sin\varphi\vec{\nabla}\varphi)T^2\\
&=-\frac{1}{g}\int_0^{2\pi}\sin\theta \sin\varphi \ d\varphi\ T^2=0,
\end{split}
\end{equation}
It means that the magnetic flux of monopoles have no contribution in directions $T^1$ and $T^2$. 
\begin{equation}
\begin{split}
&\Phi^{\text{flux}}=-\frac{1}{g}\int_0^{2\pi}r\sin\theta d\varphi\ \hat{\varphi}\cdot(1+\cos\theta)\vec{\nabla}\varphi T^3\\
&=-\frac{1}{g}(1+\cos\theta)\int_0^{2\pi}d\varphi\ T^3=-\frac{4\pi}{g}\frac{1+\cos\theta}{2} T^3,
\end{split}
\end{equation}
which illustrates the magnetic flux of the monopole attached to a Dirac string. The total magnetic flux is obtained from the contribution of monopole in direction $T^3$. 

Next, we calculate the components of field strength tensor. If we use Eqn. (\ref{gauge}) to obtain the field strength, the Yang-Mills action would not be invariant under the gauge transformation $U$. In order to get an invariant Yang-Mills action, we  use the following equation for the field strength tensor.
\begin{equation}
\begin{split}
\vec{\mathcal{F}}_{\mu\nu}^U.\vec{T}&=U\vec{F}_{\mu\nu}.\vec{T}U^{-1}=\vec{F}_{\mu\nu}^U.\vec{T}-\frac{i}{g}U[\partial_\mu,\partial_\nu]U^{-1}\\
&=(\partial_\mu\vec{A}_\nu^U-\partial_\nu \vec{A}_\mu^U+g\vec{A}_\mu^U\times \vec{A}_\nu^U).\vec{T}-\frac{i}{g}U[\partial_\mu,\partial_\nu]U^{-1}
\end{split}
\end{equation}
In fact the second term in the first line of the above equation cancels the existence of the singularity of the first term and as a result a gauge invariant action is obtained.
$\vec{\mathcal{F}}_{\mu\nu}^U$ can be written in three parts
\begin{equation}
\begin{split}
&\vec{F}_{\mu\nu}^{\text{linear}}=\partial_\mu \vec{A}_\nu^U-\partial_\nu \vec{A}_\mu^U,\\
&\vec{F}_{\mu\nu}^{\text{bilinear}}=g(\vec{A}_\mu^U\times \vec{A}_\nu^U),\\
&\vec{F}_{\mu\nu}^{\text{sing}}=-\frac{i}{g}U[\partial_\mu,\partial_\nu]U^{-1}\label{singu}
\end{split}
\end{equation}
Which agrees with what is obtained in Reference \cite{ichie} using Abelian Projection. Finally, with an Abelian projection, the contribution of the Dirac and anti-Dirac 
strings cancel each other and the monopole appears in the theory.

\begin{figure}[ht]
\centerline{\includegraphics[width=9cm]{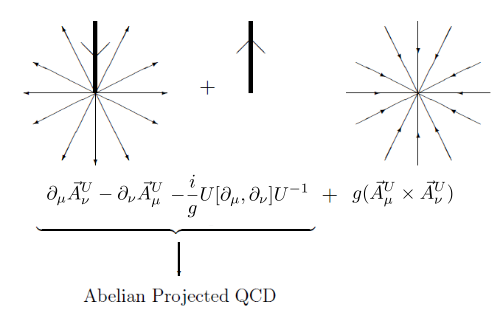}}
\caption{\label{f_munu}\small Emergence of monopoles in Abelian projected QCD\cite{ichie}}
\end{figure}

Now, using the gauge field $\vec{A}_\mu^U$ defined in Eqn. (\ref{amu}), we calculate the linear and bilinear terms of field strength tensor.
\begin{widetext}
\begin{equation}
\begin{split}
\vec{F}_{\mu\nu}^{\text{linear}}&=(\partial_\mu A_\nu^3-\partial_\nu A_\mu^3)\hat{m}+(\partial_\mu A_\nu ^1-\partial_\nu A_\mu^1)\hat{m}_1+(\partial_\mu A_\nu ^2-\partial_\nu A_\mu^2)\hat{m}_2\\
&+(A_\nu^{(m)}\partial_\mu\hat{m}-A_\mu^{(m)}\partial_\nu\hat{m})+(A_\nu^1\partial_\mu\hat{m}_1-A_\mu^1\partial_\nu\hat{m}_1)+(A_\nu^2\partial_\mu\hat{m}_2-A_\mu^2\partial_\nu\hat{m}_2)\\
&+\frac{1}{g}(\hat{m}_1.[\partial_\mu,\partial_\nu]\hat{m}_2)\hat{m}-\frac{1}{g}\hat{m}\times[\partial_\mu,\partial_\nu]\hat{m}\label{line}
\end{split}
\end{equation}
\begin{equation}
\begin{split}
\vec{F}_{\mu\nu}^{\text{bilinear}}&=g(A_\mu ^2A_\nu^3 -A_\nu ^2A_\mu^3 )\hat{m}_1+g(A_\mu^3A_\nu ^1-A_\nu^3A_\mu ^1)\hat{m}_2+g(A_\mu^1A_\nu^2-A_\nu^1A_\mu^2)\hat{m}\\
&-(A_\nu^{(m)}\partial_\mu\hat{m}-A_\mu^{(m)}\partial_\nu\hat{m})-(A_\nu^1\partial_\mu\hat{m}_1-A_\mu^1\partial_\nu\hat{m}_1)-(A_\nu^2\partial_\mu\hat{m}_2-A_\mu^2\partial_\nu\hat{m}_2)\\\label{beline}
\end{split}
\end{equation}
\end{widetext}
Using the gauge function $U$ of Eqn. (\ref{50}), the singular term of the field strength is obtained
\begin{widetext}
\begin{equation}
\begin{split}
\vec{F}_{\mu\nu}^{\text{sing}}&=(-\frac{1}{g}\sin\varphi[\partial_\mu,\partial_\nu]\theta+\frac{1}{g}\sin\theta\cos\varphi[\partial_\mu,\partial_\nu]\varphi)T^1+(\frac{1}{g}\cos\varphi[\partial_\mu,\partial_\nu]\theta+\frac{1}{g}\sin\theta\sin\varphi[\partial_\mu,\partial_\nu]\varphi)T^2\\
&+\frac{1}{g}(1+\cos \theta)[\partial_\mu,\partial_\nu]\varphi T^3\\
&=-\frac{1}{g}(\hat{m}_1.[\partial_\mu,\partial_\nu]\hat{m}_2)\hat{m}+\frac{1}{g}\hat{m}\times[\partial_\mu,\partial_\nu]\hat{m}
\end{split}
\end{equation}
\end{widetext}
We recall that $\vec{F}_{\mu\nu}^{\text{linear}}$ represents a monopole attached to a Dirac string, $\vec{F}_{\mu\nu}^{\text{bilinear}}$ describes an anti-monopole and $\vec{F}_{\mu\nu}^{\text{sing}}$ indicates an anti-Dirac string. If we add the linear, bilinear, and singular terms, the singularities are canceled and at the end, only the regular fields are remained and therefore the full QCD is restored. 
\begin{equation}
\begin{split}
\vec{F}_{\mu\nu}&=(\partial_\mu A_\nu^3-\partial_\nu A_\mu^3)\hat{m}+g(A_\mu^1A_\nu^2-A_\nu^1A_\mu^2)\hat{m}\\
&+(\partial_\mu A_\nu ^1-\partial_\nu A_\mu^1)\hat{m}_1+g(A_\mu ^2A_\nu^3 -A_\nu ^2A_\mu^3 )\hat{m}_1\\
&+(\partial_\mu A_\nu ^2-\partial_\nu A_\mu^2)\hat{m}_2+g(A_\mu^3A_\nu ^1-A_\nu^3A_\mu ^1)\hat{m}_2
\end{split}\label{full}
\end{equation}
It is also possible to obtain the linear and bilinear field strengths using the gauge field components in Eqn. (\ref{aut}) in a different way. One can easily show that 
the result is the same as what is obtained in Eqns. (\ref{line}) and (\ref{beline}).

Therefore, using a nontrivial gauge transformation $U$, a local color frame $\hat{n}_a$ is defined and it is shown that it  includes magnetic monopole. The transformed gauge field $\vec{A}_\mu^U$ in Eqn. (\ref{amu}) takes the form of Cho decomposition. After calculating the transformed field strength tensor, we find that it consists of three parts: linear, bilinear and singular. The linear part describes a monopole attached to a Dirac string, the bilinear part describes anti-monopole and the singular part describes anti-Dirac string. When we sum these three parts together, we get full QCD. This result is in agreement with what has been done in Reference \cite{ichie}.

\subsection{Agreement with the results obtained by Cho}
So far, monopoles have been extracted for SU($2$) gauge group. Monopole have already been extracted in a different way by Cho, and their condensation have been investigated for SU($2$) \cite{frank} and SU($3$) \cite{frank18} gauge groups. By imposing the maximal Abelian isometry on $\vec{A}_\mu$ which makes $\hat{m}$ a covariant constant, the restricted potential is obtained by Cho et al.
\begin{equation}
D_\mu\hat{m}=(\partial_\mu +g A_\mu\times)\hat{m}=0\Rightarrow \hat{A}_\mu=A_\mu^{(m)}\hat{m}-\frac{1}{g}\hat{m}\times\partial_\mu\hat{m}
\end{equation}
When $A_\mu^{(m)}=0$ and $\hat{m}=\hat{r}$, the restricted potential describes precisely the Wu-Yang monopole.

In this work, using a nontrivial gauge transformation U, we define a local color frame $\hat{m}_a$ that contains monopoles. Under U transformation, the gauge field is transformed to Eqn. (\ref{amu}) which is in the form of extended Cho decomposition written in the frame $\hat{m}_a$, which contains monopoles. By setting $A_\mu^1$ and $A_\mu^2$ to zero, we reach the restricted Cho decomposition with $\hat{m}=\hat{r}$ that describes magnetic monopoles.

We are looking for configurations that include monopole and vortex simultaneously (chains). In the following sections, we define another color frame $\hat{n}_a$ that contains monopole and vortex by applying another gauge transformation V on top of the previously considered monopole description. In this paper, we study SU($2$) gauge group. This way can be extended to higher gauge groups like SU($3$) with some more efforts. 
\\

We recall that $\hat{m}=\hat{r}$ corresponds to a Wu-Yang monopole. This configuration can be obtained with $\alpha=\varphi, \beta=\theta$ in Eqn. (\ref{39}), where $\theta , \varphi$ are the polar angles associated to $\hat{r}$. Since $R\hat{e}_3$ is independent of $\gamma$, any choice for $\gamma$ can be chosen.
Using Eqns. (\ref{39}) and (\ref{cmu}) and selecting $\gamma=\varphi$, we have \cite{oxman}
\begin{equation}
\begin{split}
C_\mu^{(m)}&=\frac{1}{g}(\cos \beta\partial_\mu\alpha+\partial_\mu\gamma)\bigg|_{\beta=\theta, \alpha=\gamma=\varphi}\\
&=\frac{1}{g}(\cos \theta+1)\partial_\mu\varphi,\label{Cm}
\end{split}
\end{equation}
The components of this magnetic potential are
\begin{equation}
C_r=C_\theta =0\   ,    \     C_\varphi=\frac{1}{g}\frac{1+\cos \theta}{r\sin \theta}.
\end{equation}
The above magnetic potential has a singularity at $\theta=0$. Therefore, this magnetic potential describes a Wu-Yang monopole located at the origin and attached to a Dirac string along the positive z-axis.

The choice of $\gamma$ is associated with the position of the Dirac string \cite{oxman}. If one select $\gamma=-\varphi$, magnetic potential changes to
\begin{equation}
C_\mu^{(m)}=\frac{1}{g}(\cos\theta-1)\partial_\mu\varphi.
\end{equation}
The above magnetic potential has a singularity at $\theta=\pi$. Therefore, this magnetic potential describes a Wu-Yang monopole located at the origin and attached to a Dirac string along the negative z-axis.

As explained earlier, assuming that $A_\mu^{(m)}=0$, $A_\mu^a=0$ for $a=1,2$ and also considering a hedgehog form $\hat{m}=\hat{r}$, a structure which indicates Wu-Yang monopoles in $4D$ \cite{wu} is found. Therefore by replacing $A_\mu^1=A_\mu^2=0$ and $A_\mu^3=C_\mu^{(m)}$ in Eqn. (\ref{full}), the field strength is
\begin{equation}
\vec{F}_{\mu\nu}=(\partial_\mu C_\nu^{(m)}-\partial_\nu C_\mu^{(m)})\hat{m}=F_{\mu\nu}^{(m)}\hat{m}
\end{equation}
Thus, the Lagrangian density for Wu-Yang monopoles is
\begin{equation}
\begin{split}
\mathcal{L}&=-\frac{1}{4}\vec{F}_{\mu\nu}.\vec{F}_{\mu\nu}\\
&=-\frac{1}{4}(\partial_\mu C_\nu^{(m)}-\partial_\nu C_\mu^{(m)})^2,\label{Lag}
\end{split}
\end{equation}
where $C_\mu^{(m)}$ is the magnetic potential for Wu-Yang monopoles.

The fact that the magnetic potential has a Dirac string singularity is an undesirable feature for Lagrangian (\ref{Lag}). In order to provide a field theoretical description of the theory, we have to remove the Dirac string singularity. To do this, we consider the dual field strength tensor $F_{\mu\nu}^{*(m)}=\frac{1}{2}\epsilon_{\mu\nu\rho\sigma}F^{\rho\sigma(m)}$ that can be described by a regular potential with no Dirac string singularity \cite{cho1}. So the dual magnetic potential $C_\mu^{*(m)}$ is defined by
\begin{equation}
F^{*(m)}_{\mu\nu}=\partial_\mu C_\mu^{*(m)}-\partial_\nu C_\mu^{*(m)}.
\end{equation}
$C_\mu^{*(m)}$ can describe the monopole and does not contain the Dirac string singularity anymore. Now we can replace $C_\mu^{(m)}$ in favor of $C_\mu^{*(m)}$ in the Lagrangian (\ref{Lag}).

For a field theoretical description of the monopole, a new field for the monopole should be introduced. Since a monopole is a point-like object and does not have any obvious spin structure, it may be described by a complex scalar field $\phi$. Naturally we would expect that $\phi$ should couple to $C_\mu^{*(m)}$ minimally. By adding a potential and a kinetic term for $\phi$ \cite{cho1}, the Lagrangian for monopoles is completed.  
\begin{equation}
\begin{split}
\mathcal{L}=&-\frac{1}{4}(\partial_\mu C_\nu^{*(m)}-\partial_\nu C_\mu^{*(m)})^2\\
&+|(\partial_\mu+igC_\mu^{*(m)})\phi|^2-\frac{m^2}{2}(\phi\phi^*)-\lambda(\phi\phi^*)^2.
\end{split}
\end{equation}
By adding the scalar potential, the spontaneous symmetry breaking would be possible and one can examine the condensation of monopoles and confinement.
\\

\section{Center vortices as defects of the local color frame}\label{sec4}

Center vortices are color magnetic line-like (surface-like) defects in three (four) dimensions. Their quantization is done in terms of center elements of the gauge group. When a center vortex is linked by a Wilson loop, the Wilson loop variable gets an element of the center $Z(N)$.  

To observe vortices in the theory, it is not necessary to use a nontrivial gauge transformation as it is done for monopoles in Eqn. (\ref{a}). Rather, by proposing the following configuration, closed thin center vortices are introduced
\begin{equation}
\vec{A}_\mu^{\textsl{thin}}.\vec{T}=VA_\mu^aT^aV^{-1}+\frac{i}{g}V\partial_\mu V^{-1}-\text{ideal vortex}\label{ideal}
\end{equation}
where $V\in SU(N)$. The ideal vortex is localized on the three-volume where the transformation $V$ is discontinuous \cite{engel, rein}.

For example for the SU($2$) gauge group, the gauge transformation can be parametrized by $V=e^{i\varphi T_3}$ and we introduce a local basis $\hat{n}'_a$ in the color space
\begin{equation}
VT^aV^{-1}=\hat{n}'_a.\vec{T}\   \   ,   \   \hat{n}'_a=R_3\hat{e}_a, \label{S-Gauge}
\end{equation}
\begin{equation}
V=e^{i\varphi T_3}=\left(
\begin{array}{cc}
e^{i\frac{\varphi}{2}}&0\\
0&e^{-i\frac{\varphi}{2}}
\end{array}
\right).
\end{equation}
Using matrix multiplication, the components of the color frame $\hat{n}'$ can be obtained as follows
\begin{equation}
VT^1V^{-1}=\hat{n}'_1.\vec{T}\Rightarrow \hat{n}'_1=
\begin{pmatrix}
\cos\varphi\\
-\sin\varphi\\
0
\end{pmatrix}
\end{equation}
\begin{equation}
VT^2V^{-1}=\hat{n}'_2.\vec{T}\Rightarrow \hat{n}'_2=
\begin{pmatrix}
\sin\varphi\\
\cos\varphi\\
0
\end{pmatrix}
\end{equation}
\begin{equation}
VT^3V^{-1}=\hat{n}'_3.\vec{T}\Rightarrow \hat{n}'_3=
\begin{pmatrix}
0\\
0\\
1
\end{pmatrix}
\end{equation}
Using the definition of (\ref{cmu}) and the above definitions for $\hat{n}'_1$ and $\hat{n}'_2$, we calculate the magnetic potential of vortices
\begin{equation}
C_\mu^{(v)}=-\frac{1}{g}\hat{n}'_1.\partial_\mu\hat{n}'_2=-\frac{1}{g}\partial_\mu\varphi
\end{equation}
\begin{equation}
C_\mu^{(v)}=-\frac{1}{g}\partial_\mu\varphi\rightarrow C_\rho =C_z=0\   \    ,     \    C_\varphi=-\frac{1}{g}\frac{1}{\rho}
\end{equation}
where $\rho$ is the distance from the $z$-axis in the cylindrical coordinate system.

The mapping $V$ is not single valued and
\begin{equation}
\frac{i}{g}e^{i\varphi T_3}\partial_\mu e^{-i\varphi T_3}=\frac{1}{g}\partial_\mu\varphi\delta^{a3}T^a+\text{ideal vortex}.\label{thin}
\end{equation}
Replacing Eqn. (\ref{thin}) and $VA_\mu^aT^aV^{-1}=A_\mu^a\hat{n}'_a.\vec{T}$ in Eqn. (\ref{ideal}) leads to the following representation for thin center vortices \cite{thinv}
\begin{equation}
\vec{A}_\mu^{\textsl{V}}.\vec{T}=\bigg((\frac{1}{g}\partial_\mu\varphi+A_\mu^3)\hat{n}'_3+A_\mu^1\hat{n}'_1+A_\mu^2\hat{n}'_2\bigg).\vec{T}. \label{A-nprime}
\end{equation}

Using the above equation and setting $\hat{n}'=\hat{n}'_3$, we calculate the field strength tensor. According to what was mentioned in the previous sections about the effect of gauge transformations on Lagrangian invariance, if we define the field strength tensor as follows, the Lagrangian does not remain invariant under gauge transformations 
\begin{equation}
\begin{split}
\vec{F}_{\mu\nu}^{V}&=\partial_\mu \vec{A}_\nu^V-\partial_\nu \vec{A}_\mu^V+g \vec{A}_\mu^V\times \vec{A}_\nu^V\\
&=(\partial_\mu A_\nu^3-\partial_\nu A_\mu^3)\hat{n}'+g(A_\mu^1 A_\nu^2- A_\mu^2 A_\nu^1)\hat{n}'\\
&+(\partial_\mu A_\nu^1- \partial_\nu A_\mu^1)\hat{n}'_1 +g(A_\nu^3A_\mu^2-A_\mu^3A_\nu^2)\hat{n}'_1\\
&+(\partial_\mu A_\nu^2-\partial_\nu A_\mu^2)\hat{n}'_2+g(A_\mu^3A_\nu^1-A_\nu^3A_\mu^1)\hat{n}'_2\\
&+\frac{1}{g}[\partial_\mu.\partial_\nu]\varphi\hat{n}',\label{vortexx}
\end{split}
\end{equation}
where the last term describes the vortex. But if we use the following equation to calculate the field strength tensor, Lagrangian will be invariant under gauge transformations.
\begin{equation}
\begin{split}
\vec{F}_{\mu\nu}^V.&\vec{T}=V\vec{F}_{\mu\nu}.\vec{T}V^{-1}\\
&=(\partial_\mu \vec{A}_\nu^V-\partial_\nu \vec{A}_\mu^V+g \vec{A}_\mu^V\times \vec{A}_\nu^V).\vec{T}-\frac{i}{g}V[\partial_\mu.\partial_\nu]V^{-1}
\end{split}
\end{equation}
where,
\begin{equation}
-\frac{i}{g}V[\partial_\mu.\partial_\nu]V^{-1}=-\frac{1}{g}[\partial_\mu.\partial_\nu]\varphi T^3=-\frac{1}{g}[\partial_\mu.\partial_\nu]\varphi\hat{n}'
\end{equation}
Therefore
\begin{equation}
\begin{split}
\vec{F}_{\mu\nu}^V&=(\partial_\mu A_\nu^3-\partial_\nu A_\mu^3)\hat{n}'+g(A_\mu^1 A_\nu^2- A_\mu^2 A_\nu^1)\hat{n}'\\
&+(\partial_\mu A_\nu^1- \partial_\nu A_\mu^1)\hat{n}'_1 +g(A_\nu^3A_\mu^2-A_\mu^3A_\nu^2)\hat{n}'_1\\
&+(\partial_\mu A_\nu^2-\partial_\nu A_\mu^2)\hat{n}'_2+g(A_\mu^3A_\nu^1-A_\nu^3A_\mu^1)\hat{n}'_2,
\end{split}\label{vort}
\end{equation}
which includes regular gauge fields and describes the full QCD.

Now we use Eqn. (\ref{vortexx}) and calculate the Lagrangian density for describing center vortices
\begin{equation}
\begin{split}
\mathcal{L}&=-\frac{1}{4}\vec{F}_{\mu\nu}^V.\vec{F}_{\mu\nu}^V\\
&=-\frac{1}{4}\big(\partial_\mu (A_\nu^3+\frac{1}{g}\partial_\nu\varphi)-\partial_\nu (A_\mu^3+\frac{1}{g}\partial_\mu\varphi)\big)^2\\
&-\frac{1}{4}(\partial_\mu A_\nu^1- \partial_\nu A_\mu^1)^2-\frac{1}{4}(\partial_\mu A_\nu^2-\partial_\nu A_\mu^2)^2\\
&-\frac{1}{4}g^2\bigg((A_\mu^1 A_\nu^2- A_\mu^2 A_\nu^1)^2+(A_\nu^3A_\mu^2-A_\mu^3A_\nu^2)^2\\
&+(A_\mu^3A_\nu^1-A_\nu^3A_\mu^1)^2\bigg)
\end{split}
\end{equation}
The first term illustrates the kinetic energy of center vortices, and the other terms illustrate the kinetic energy and interaction between regular fields. By adding proper scalar field for center vortices as Higgs field, one can investigate the spontaneous symmetry breaking and condensation of vortices. Since the vortices are line-like objects, their condensation needs further investigation.

\section{Correlated monopoles and vortices}\label{sec3}

In this section, it is shown that by using Cho decomposition method and parameterizing the gauge fields appropriately, structures involving vortices attached to monopoles, called  chains, appear in the theory. We use an extended class of frames $\hat{n}_a$, obtained by introducing a $V$-sector on top of the previously considered monopole description \cite{oxman}. Then, by extending $U\rightarrow VU$, the frame $\hat{n}_a, a=1,2,3$ is defined, where $U$ is the same transformation used in Section \ref{sec2} for describing the monopoles.
\begin{equation}
(VU)T^a(VU)^{-1}=VUT^aU^{-1}V^{-1}=\hat{n}_a.\vec{T},
\end{equation}
\begin{equation}
\hat{n}_a=R(VU)\hat{e}_a=R(V)R(U)\hat{e}_a=R(V)\hat{m}_a.
\end{equation}

$V\in SU(2)$ is not single valued along any closed loop. For instance, we can choose $V$ including a rotation that leaves $\hat{m}$ invariant
\begin{equation}
V=e^{-i\xi \hat{m}.\vec{T}},
\end{equation}
where if $\xi$ changes by $2\pi$, a rotation around the vortex is understood. We can also write
\begin{equation}
VU=e^{-i\xi UT_3U^{-1}}U=UV_3   \   \   ,   \   V_3=e^{-i\xi T_3}.
\end{equation}
\\
With this parametrization, the frame $\hat{n}_a$ changes as the following
\begin{equation}
\hat{n}_a=R_m(\xi)\hat{m}_a\   \   ,   \   R_m(\xi)=e^{\xi \hat{m}.\vec{M}}\   \   ,   \    \hat{n}=\hat{m}.\label{frame}
\end{equation}
Applying the definition of $\hat{n}_a$ to $C_\mu^{(n)}=-\frac{1}{g}\hat{n}_1.\partial_\mu\hat{n}_2$ in Eqn. (\ref{cmu}), the magnetic potential changes as follows
\begin{equation}
C_\mu^{(n)}=-\frac{1}{g}R_m\hat{m}_1.[R_m(\partial_\mu\hat{m}_2)+(\partial_\mu R_m)\hat{m}_2]=C_\mu^{(m)}+C_\mu^{(v)},\label{magnet}
\end{equation}
where
\begin{equation}
\begin{split}
C_\mu^{(m)}&=-\frac{1}{g}\hat{m}_1.\partial_\mu\hat{m}_2,\\
C_\mu^{(v)}&=-\frac{1}{g}\hat{m}_1.(R_m^{-1}\partial_\mu R_m)\hat{m}_2=\frac{1}{g}\partial_\mu\xi.
\end{split}
\end{equation}
\\
The magnetic potential $C_\mu^{(n)}$ in Eqn. (\ref{magnet}) is written as a sum of two parts: magnetic potentials of monopoles and vortices. $C_\mu^{(v)}$ represents the potential of a center vortex in the outside of the core without any additional ideal vortex, while $C_\mu^{(m)}$ represents the magnetic potential of a monopole discussed in the Section \ref{sec2}. Therefore, by introducing a local frame constructed by $VU$ transformation, the relationship between monopoles and vortices is determined. That is, the vortex surfaces are connected by the monopole worldlines. 
\\

Using Eqn. (\ref{50}) and selecting $\xi=-\varphi$, $VU$ is obtained as follows
\begin{equation}
VU=\left(
\begin{array}{cc}
\cos\frac{\theta}{2}e^{-i\frac{\varphi}{2}}&-\sin\frac{\theta}{2}e^{-i\frac{\varphi}{2}}\\
\sin\frac{\theta}{2}e^{i\frac{\varphi}{2}}&\cos\frac{\theta}{2}e^{i\frac{\varphi}{2}}
\end{array}
\right)
\end{equation}
Therefore the components of the local frame $\hat{n}_a$ is obtained as follows
\begin{equation}
(VU)T^1(VU)^{-1}=\hat{n}_1.\vec{T}\Rightarrow \hat{n}_1=
\begin{pmatrix}
\cos\theta\cos\varphi\\
\cos\theta\sin\varphi\\
-\sin\theta
\end{pmatrix}=\hat{\theta}\label{n1}
\end{equation}
\begin{equation}
(VU)T^2(VU)^{-1}=\hat{n}_2.\vec{T}\Rightarrow \hat{n}_2=
\begin{pmatrix}
-\sin\varphi\\
\cos\varphi\\
0
\end{pmatrix}=\hat{\varphi}\label{n2}
\end{equation}
\begin{equation}
(VU)T^3(VU)^{-1}=\hat{n}.\vec{T}\Rightarrow \hat{n}=
\begin{pmatrix}
\sin\theta\cos\varphi\\
\sin\theta\sin\varphi\\
\cos\theta\label{n}
\end{pmatrix}=\hat{m}=\hat{r}
\end{equation}
\\
Using Eqn. (\ref{cmum})
\begin{equation}
C_\mu^{(m)}=\frac{1}{g}(1+\cos \theta)\partial_\mu\varphi,
\end{equation}
and selecting $\xi=-\varphi$
\begin{equation}
C_\mu^{(v)}=\frac{1}{g}\partial_\mu\xi\bigg|_{\xi=-\varphi}=-\frac{1}{g}\partial_\mu\varphi,
\end{equation}
we have
\begin{equation}
C_\mu^{(n)}=C_\mu^{(m)}+C_\mu^{(v)}=\frac{1}{g}\cos\theta\partial_\mu\varphi.\label{cmn}
\end{equation}
Eqn. (\ref{cmn}) can also be obtained directly using $C_\mu^{(n)}=-\frac{1}{g}\hat{n}_1.\partial_\mu\hat{n}_2$ and using the $\hat{n}_1$ and $\hat{n}_2$ components in Eqns. (\ref{n1}) and (\ref{n2}).

Eqn. (\ref{cmn}) becomes non-zero for $\theta=0$ and $\pi$. It means that there are some defects on the positive and negative z-axis. Since this potential is of vortex type potential, Eqn. (\ref{cmn}) represents two vortex-line, one in the positive z-axis and the other one in the negative z-axis, and they are connected by a monopole located at the origin \cite{oxman}
\\

Now we continue the work done in reference \cite{oxman} to obtain an explicit Lagrangian for the defects and their interactions. By calculating the field strength tensor for a configuration involving monopoles and vortices, we obtain a Lagrangian density which clearly illustrates the interaction between monopoles and vortices.\\

Under $VU$ gauge transformation, the gauge field is transformed as follows
\begin{equation}
\begin{split}
&\vec{A}_\mu^{VU}.\vec{T}=(VU)\vec{A}_\mu.\vec{T}(VU)^{-1}+\frac{i}{g}(VU)\partial_\mu(VU)^{-1}\\
&=\bigg((A_\mu^3-C_\mu^{(m)}-C_\mu^{(v)})\hat{n}-\frac{1}{g}\hat{n}\times\partial_\mu\hat{n}+A_\mu^1\hat{n}_1+A_\mu^2\hat{n}_2\bigg).\vec{T}\\
&=(A_\mu^{VU})^1T^1+(A_\mu^{VU})^2T^2+(A_\mu^{VU})^3T^3,
\end{split}
\end{equation}
where
\begin{widetext}
\begin{equation}
\begin{split}
(A_\mu^{VU})^1&=\cos\theta\cos\varphi A_\mu^1-\sin\varphi A_\mu^2+\sin\theta\cos\varphi A_\mu^3+\frac{1}{g}\sin\varphi\partial_\mu\theta\\
(A_\mu^{VU})^2&=\cos\theta\sin\varphi A_\mu^1+\cos\varphi A_\mu^2+\sin\theta\sin\varphi A_\mu^3-\frac{1}{g}\cos\varphi\partial_\mu\theta\\
(A_\mu^{VU})^3&=-\sin\theta A_\mu^1+\cos\theta A_\mu^3-\frac{1}{g}\partial_\mu\varphi\label{AVU}
\end{split}
\end{equation}
\end{widetext}
The first three terms of $(A_\mu^{VU})^1$ and $(A_\mu^{VU})^2$ include regular fields and we would like to interpret their last terms
\begin{equation}
\begin{split}
\frac{1}{g}\sin\varphi\vec{\nabla}\theta&=\frac{1}{g}\frac{1}{r}\sin\varphi\ \hat{\theta}\\
-\frac{1}{g}\cos\varphi\vec{\nabla}\theta&=-\frac{1}{g}\frac{1}{r}\cos\varphi\ \hat{\theta}.
\end{split}
\end{equation}
\\
These terms are proportional to $\frac{1}{r}$ and describe contribution of a monopole at the origin. The first two terms of $(A_\mu^{VU})^3$ include regular fields and the last term describes contribution of monopole and center vortex. 

Now we calculate all terms of field strength tensor and interpret them. Similar to what is mentioned in the previous sections, in order to have an invariant Lagrangian under gauge transformations, we use the following equation to obtain the field strength tensor
\begin{widetext}
\begin{equation}
\begin{split}
\vec{\mathcal{F}}_{\mu\nu}^{VU}.\vec{T}&=(VU)\vec{F}_{\mu\nu}(VU)^{-1}\\
&=(\partial_\mu \vec{A}_\nu^{VU}-\partial_\nu \vec{A}_\mu^{VU}+g \vec{A}_\mu^{VU}\times \vec{A}_\nu^{VU}).\vec{T}-\frac{i}{g}(VU)[\partial_\mu.\partial_\nu](VU)^{-1},
\end{split}
\end{equation}
\end{widetext}
Which can be written as a sum of three sentences: linear, bilinear and singular.
\begin{widetext}
\begin{equation}
\begin{split}
\vec{F}_{\mu\nu}^{\text{linear}}&=\partial_\mu \vec{A}_\nu^{VU}-\partial_\nu \vec{A}_\mu^{VU}\\
&-(\partial_\mu A_\nu^3-\partial_\nu A_\mu^3)\hat{n}+(\partial_\mu A_\nu ^1-\partial_\nu A_\mu^1)\hat{n}_1+(\partial_\mu A_\nu ^2-\partial_\nu A_\mu^2)\hat{n}_2\\
&+(A_\nu^{(n)}\partial_\mu\hat{n}-A_\mu^{(n)}\partial_\nu\hat{n})+(A_\nu^1\partial_\mu\hat{n}_1-A_\mu^1\partial_\nu\hat{n}_1)+(A_\nu^2\partial_\mu\hat{n}_2-A_\mu^2\partial_\nu\hat{n}_2)\\
&+\frac{1}{g}(\hat{n}_1.[\partial_\mu,\partial_\nu]\hat{n}_2)\hat{n}-\frac{1}{g}\hat{n}\times[\partial_\mu,\partial_\nu]\hat{n}
\end{split}
\end{equation}
\begin{equation}
\begin{split}
\vec{F}_{\mu\nu}^{\text{bilinear}}&=g \vec{A}_\mu^{VU}\times \vec{A}_\nu^{VU}\\
&=g(A_\mu ^2A_\nu^3 -A_\nu ^2A_\mu^3 )\hat{n}_1+g(A_\mu^3A_\nu ^1-A_\nu^3A_\mu ^1)\hat{n}_2+g(A_\mu^1A_\nu^2-A_\nu^1A_\mu^2)\hat{n}\\
&-(A_\nu^{(n)}\partial_\mu\hat{n}-A_\mu^{(n)}\partial_\nu\hat{n})-(A_\nu^1\partial_\mu\hat{n}_1-A_\mu^1\partial_\nu\hat{n}_1)-(A_\nu^2\partial_\mu\hat{n}_2-A_\mu^2\partial_\nu\hat{n}_2)
\end{split}
\end{equation}
\begin{equation}
\begin{split}
\vec{F}_{\mu\nu}^{\text{sing}}&=-\frac{i}{g}(VU)[\partial_\mu.\partial_\nu](VU)^{-1}\\
&=-\frac{1}{g}\sin\varphi[\partial_\mu.\partial_\nu]\theta T^1+\frac{1}{g}\cos\varphi[\partial_\mu.\partial_\nu]\theta T^2+\frac{1}{g}[\partial_\mu.\partial_\nu]\varphi T^3\\
&=-\frac{1}{g}(\hat{n}_1.[\partial_\mu,\partial_\nu]\hat{n}_2)\hat{n}+\frac{1}{g}\hat{n}\times[\partial_\mu,\partial_\nu]\hat{n},\label{Fsingu}
\end{split}
\end{equation}
\end{widetext}
where $A_\mu^{(n)}=A_\mu^3-C_\mu^{(m)}-C_\mu^{(v)}$. $\vec{F}_{\mu\nu}^{\text{linear}}$ describes the monopole and vortex, $\vec{F}_{\mu\nu}^{\text{bilinear}}$ describes an anti-monopole and $\vec{F}_{\mu\nu}^{\text{sing}}$ describes an anti-vortex. However, if we add the linear, bilinear, and singular field strengths together, eventually the regular fields remain which describe full QCD.
\begin{equation}
\begin{split}
\vec{F}_{\mu\nu}&=(\partial_\mu A_\nu^3-\partial_\nu A_\mu^3)\hat{n}+g(A_\mu^1A_\nu^2-A_\nu^1A_\mu^2)\hat{n}\\
&+(\partial_\mu A_\nu ^1-\partial_\nu A_\mu^1)\hat{n}_1+g(A_\mu ^2A_\nu^3 -A_\nu ^2A_\mu^3 )\hat{n}_1\\
&+(\partial_\mu A_\nu ^2-\partial_\nu A_\mu^2)\hat{n}_2+g(A_\mu^3A_\nu ^1-A_\nu^3A_\mu ^1)\hat{n}_2
\end{split}
\end{equation}
If we select $A_\mu^1=A_\mu^2=0$ and $A_\mu^3=C_\mu^{(m)}+C_\mu^{(v)}$
\begin{equation}
\vec{F}_{\mu\nu}=\big(\partial_\mu(C_\nu^{(m)}+C_\nu^{(v)})-\partial_\nu(C_\mu^{(m)}+C_\mu^{(v)})\big)\hat{n}
\end{equation}
According to Eqn. (\ref{cmn}),
\begin{equation}
C_\mu^{(m)}+C_\mu^{(v)}=\frac{1}{g}\cos\theta\partial_\mu\varphi\Rightarrow C_r=C_\theta=0, C_\varphi=\frac{1}{g}\frac{\cos\theta}{r\sin\theta}
\end{equation}
$C_\varphi$ diverges for $r=0$, $\theta=0$ and $\theta=\pi$. Therefore, there are some defects at the origin, positive and negative z-axis that include a monopole at the origin and two vortex-line on positive and negative z-axis which form a chain. We can write the Lagrangian density for this configuration
\begin{equation}
\begin{split}
\mathcal{L}&=-\frac{1}{4}\vec{F}_{\mu\nu}.\vec{F}_{\mu\nu}\\
&=-\frac{1}{4}\big(\partial_\mu(C_\nu^{(m)}+C_\nu^{(v)})-\partial_\nu(C_\mu^{(m)}+C_\mu^{(v)})\big)^2\\
&=-\frac{1}{4}(\partial_\mu C_\nu^{(m)}-\partial_\nu C_\mu^{(m)})^2-\frac{1}{4}(\partial_\mu C_\nu^{(v)}-\partial_\nu C_\mu^{(v)})^2\\
&-\frac{1}{2}(\partial_\mu C_\nu^{(m)}-\partial_\nu C_\mu^{(m)})(\partial_\mu C_\nu^{(v)}-\partial_\nu C_\mu^{(v)})
\end{split}
\end{equation}
The above Lagrangian describes correlated monopole and vortices. The first term describes kinetic energy of monopole, the second term describes kinetic energy of vortices and the third term describes the interaction between monopole and vortices. By adding proper scalar fields for monopole and vortices as Higgs fields, the spontaneous symmetry breaking and  condensations can be investigated. 
The components of $F_{\mu\nu}^{\text{linear}}$ and $F_{\mu\nu}^{\text{bilinear}}$ are calculated using $A_\mu^{VU}$ components. We present these components in Appendix \ref{appp}.

\section{Conclusions}\label{sec5}

Even though monopoles and vortices as two of the successful candidates for QCD magnetic defects, have been able to describe some of the confinement properties, however neither of them present a complete description of confinement properties and both have shown some shortcomings in explaining some other properties of confinement potentials. Correlated monopoles and center vortices observed in lattice simulations may be the key to the solution of these shortcomings. A theory which reconciles these two defects may correctly predict the confining potential behaviors including N-ality dependence at large distances and Casimir scaling at intermediate distances.

In this paper, we use the Cho decomposition method \cite{cho1, cho2, cho3} and we apply the successive gauge  transformations proposed by Oxman \cite{oxman} to observe the monopole vortex junctions. The correlated monopoles and vortices can be constructed in terms of a local frame made by the two gauge transformations, such that the vortex worldsheets is connected by the Monopole worldlines. A nontrivial gauge transformation that leads to the appearance of monopoles, and an SU($2$) gauge transformation that is single valued around each closed loop leads to appearance of center vortices. Next, by comparison with Abelian Projection scenario which has been discussed in reference \cite{ichie}, a Lagrangian density for QCD vacuum in the confining regime is obtained. The Lagrangian includes kinetic energy of monopoles and vortices as well as a contribution which describes the interaction between monopole and vortices.

\appendix

\section{calculating the components of $F_{\mu\nu}^{\text{linear}}$ and $F_{\mu\nu}^{\text{bilinear}}$ for correlated monopoles and vortices}\label{appp}
In this Appendix using the components of $A_\mu^{VU}$ in Eqn. (\ref{AVU}), linear and bilinear field strengths are calculated. Finally we show that summing the linear, bilinear and singular field strengths leads to full QCD.

\begin{widetext}
\begin{equation}
\begin{split}
(F_{\mu\nu}^{\text{linear}})^1&=\partial_\mu (A_\nu^{VU})^1-\partial_\nu (A_\mu^{VU})^1\\
&=\cos\theta\cos\varphi(\partial_\mu A_\nu^1-\partial_\nu A_\mu^1)-\sin\varphi(\partial_\mu A_\nu^2-\partial_\nu A_\mu^2)+\sin\theta\cos\varphi(\partial_\mu A_\nu^3-\partial_\nu A_\mu^3)\\
&-\sin\theta\cos\varphi(A_\nu^1\partial_\mu\theta-A_\mu^1\partial_\nu\theta)-\cos\theta\sin\varphi(A_\nu^1\partial_\mu\varphi-A_\mu^1\partial_\nu\varphi)-\cos\varphi(A_\nu^2\partial_\mu\varphi-A_\mu^2\partial_\nu\varphi)\\
&+\cos\theta\cos\varphi(A_\nu^3\partial_\mu\theta-A_\mu^3\partial_\nu\theta)-\sin\theta\sin\varphi(A_\nu^3\partial_\mu\varphi-A_\mu^3\partial_\nu\varphi)\\
&+\frac{1}{g}\cos\varphi(\partial_\mu\varphi \partial_\nu\theta -\partial_\nu\varphi\partial_\mu\theta)+\frac{1}{g}\sin\varphi [\partial_\mu,\partial_\nu]\theta
\end{split}
\end{equation}
\begin{equation}
\begin{split}
(F_{\mu\nu}^{\text{linear}})^2&=\partial_\mu (A_\nu^{VU})^2-\partial_\nu (A_\mu^{VU})^2\\
&=\cos\theta\sin\varphi(\partial_\mu A_\nu^1-\partial_\nu A_\mu^1)+\cos\varphi(\partial_\mu A_\nu^2-\partial_\nu A_\mu^2)+\sin\theta\sin\varphi(\partial_\mu A_\nu^3-\partial_\nu A_\mu^3)\\
&-\sin\theta\sin\varphi(A_\nu^1\partial_\mu\theta-A_\mu^1\partial_\nu\theta)+\cos\theta\cos\varphi(A_\nu^1\partial_\mu\varphi-A_\mu^1\partial_\nu\varphi)-\sin\varphi(A_\nu^2\partial_\mu\varphi-A_\mu^2\partial_\nu\varphi)\\
&+\cos\theta\sin\varphi(A_\nu^3\partial_\mu\theta-A_\mu^3\partial_\nu\theta)+\sin\theta\cos\varphi(A_\nu^3\partial_\mu\varphi-A_\mu^3\partial_\nu\varphi)\\
&+\frac{1}{g}\sin\varphi(\partial_\mu\varphi \partial_\nu\theta -\partial_\nu\varphi\partial_\mu\theta)-\frac{1}{g}\cos\varphi [\partial_\mu,\partial_\nu]\theta
\end{split}
\end{equation}
\begin{equation}
\begin{split}
(F_{\mu\nu}^{\text{linear}})^3&=\partial_\mu (A_\nu^{VU})^3-\partial_\nu (A_\mu^{VU})^3\\
&=-\sin\theta(\partial_\mu A_\nu^1-\partial_\nu A_\mu^1)+\cos\theta(\partial_\mu A_\nu^3-\partial_\nu A_\mu^3)-\cos\theta(A_\nu^1\partial_\mu\theta-A_\mu^1\partial_\nu\theta)-\sin\theta(A_\nu^3\partial_\mu\theta-A_\mu^3\partial_\nu\theta)\\
&-\frac{1}{g}[\partial_\mu,\partial_\nu]\varphi
\end{split}
\end{equation}
\begin{equation}
\begin{split}
(F_{\mu\nu}^{\text{bilinear}})^1&=g\big((A_\mu^{VU})^2(A_\nu^{VU})^3-(A_\nu^{VU})^2(A_\mu^{VU})^3\big)\\
&=g\bigg(\cos\varphi\cos\theta(A_\mu^2 A_\nu^3-A_\nu^2A_\mu^3)-\sin\varphi(A_\mu^3A_\nu^1-A_\nu^3A_\mu^1)+\sin\theta\cos\varphi(A_\mu^1A_\nu^2-A_\nu^1A_\mu^2)\bigg)\\
&+\sin\theta\cos\varphi(A_\nu^1\partial_\mu\theta-A_\mu^1\partial_\nu\theta)+\cos\theta\sin\varphi(A_\nu^1\partial_\mu\varphi-A_\mu^1\partial_\nu\varphi)+\cos\varphi(A_\nu^2\partial_\mu\varphi-A_\mu^2\partial_\nu\varphi)\\
&-\cos\theta\cos\varphi(A_\nu^3\partial_\mu\theta-A_\mu^3\partial_\nu\theta)+\sin\theta\sin\varphi(A_\nu^3\partial_\mu\varphi-A_\mu^3\partial_\nu\varphi)-\frac{1}{g}\cos\varphi(\partial_\mu\varphi \partial_\nu\theta -\partial_\nu\varphi\partial_\mu\theta)
\end{split}
\end{equation}
\begin{equation}
\begin{split}
(F_{\mu\nu}^{\text{bilinear}})^2&=g\big((A_\mu^{VU})^3(A_\nu^{VU})^1-(A_\nu^{VU})^3(A_\mu^{VU})^1\big)\\
&=g\bigg(\sin\varphi\cos\theta(A_\mu^2 A_\nu^3-A_\nu^2A_\mu^3)+\cos\varphi(A_\mu^3A_\nu^1-A_\nu^3A_\mu^1)+\sin\theta\sin\varphi(A_\mu^1A_\nu^2-A_\nu^1A_\mu^2)\bigg)\\
&+\sin\theta\sin\varphi(A_\nu^1\partial_\mu\theta-A_\mu^1\partial_\nu\theta)-\cos\theta\cos\varphi(A_\nu^1\partial_\mu\varphi-A_\mu^1\partial_\nu\varphi)+\sin\varphi(A_\nu^2\partial_\mu\varphi-A_\mu^2\partial_\nu\varphi)\\
&-\cos\theta\sin\varphi(A_\nu^3\partial_\mu\theta-A_\mu^3\partial_\nu\theta)-\sin\theta\cos\varphi(A_\nu^3\partial_\mu\varphi-A_\mu^3\partial_\nu\varphi)-\frac{1}{g}\sin\varphi(\partial_\mu\varphi \partial_\nu\theta -\partial_\nu\varphi\partial_\mu\theta)
\end{split}
\end{equation}
\begin{equation}
\begin{split}
(F_{\mu\nu}^{\text{bilinear}})^3&=g\big((A_\mu^{VU})^1(A_\nu^{VU})^2-(A_\nu^{VU})^1(A_\mu^{VU})^2\big)\\
&=g\bigg(-\sin\theta(A_\mu^2 A_\nu^3-A_\nu^2A_\mu^3)+\cos\theta(A_\mu^1A_\nu^2-A_\nu^1A_\mu^2)\bigg)+\cos\theta(A_\nu^1\partial_\mu\theta-A_\mu^1\partial_\nu\theta)+\sin\theta(A_\nu^3\partial_\mu\theta-A_\mu^3\partial_\nu\theta)
\end{split}
\end{equation}
\end{widetext}
Now if we sum $F_{\mu\nu}^{\text{linear}}$, $F_{\mu\nu}^{\text{bilinear}}$ in the above equations and $F_{\mu\nu}^{\text{sing}}$ in Eqn. (\ref{Fsingu}), we have
\begin{widetext}
\begin{equation}
\begin{split}
F_{\mu\nu}&=F_{\mu\nu}^{\text{linear}}+F_{\mu\nu}^{\text{bilinear}}+F_{\mu\nu}^{\text{sing}}\\
&=\big(\cos\theta\cos\varphi(\partial_\mu A_\nu^1-\partial_\nu A_\mu^1)-\sin\varphi(\partial_\mu A_\nu^2-\partial_\nu A_\mu^2)+\sin\theta\cos\varphi(\partial_\mu A_\nu^3-\partial_\nu A_\mu^3)\big)T^1\\
&+\big(\cos\theta\sin\varphi(\partial_\mu A_\nu^1-\partial_\nu A_\mu^1)+\cos\varphi(\partial_\mu A_\nu^2-\partial_\nu A_\mu^2)+\sin\theta\sin\varphi(\partial_\mu A_\nu^3-\partial_\nu A_\mu^3)\big)T^2\\
&+\big(-\sin\theta(\partial_\mu A_\nu^1-\partial_\nu A_\mu^1)+\cos\theta(\partial_\mu A_\nu^3-\partial_\nu A_\mu^3)\big)T^3\\
&+g\big(\cos\theta\cos\varphi(A_\mu^2A_\nu^3-A_\mu^3A_\nu^2)-\sin\varphi(A_\mu^3A_\nu^1-A_\mu^1A_\nu^3)+\sin\theta\cos\varphi(A_\mu^1A_\nu^2-A_\nu^1A_\mu^2)\big)T^1\\
&+g\big(\cos\theta\sin\varphi(A_\mu^2A_\nu^3-A_\mu^3A_\nu^2)+\cos\varphi(A_\mu^3A_\nu^1-A_\mu^1A_\nu^3)+\sin\theta\sin\varphi(A_\mu^1A_\nu^2-A_\nu^1A_\mu^2)\big)T^2\\
&+g\big(-\sin\theta(A_\mu^2A_\nu^3-A_\mu^3A_\nu^2)+\cos\theta(A_\mu^1A_\nu^2-A_\nu^1A_\mu^2)\big)T^3\\
\end{split}
\end{equation}
\end{widetext}
After summing linear, bilinear and singular field strengths, only regular gauge fields remain and this shows full QCD. One can use the definitions of $\hat{n}_1$, $\hat{n}_2$ and $\hat{n}$ in Eqns. (\ref{n1}), (\ref{n2}) and (\ref{n}) and rewrite the above equation as follows
\begin{equation}
\begin{split}
\vec{F}_{\mu\nu}&=(\partial_\mu A_\nu^3-\partial_\nu A_\mu^3)\hat{n}+g(A_\mu^1A_\nu^2-A_\nu^1A_\mu^2)\hat{n}\\
&+(\partial_\mu A_\nu ^1-\partial_\nu A_\mu^1)\hat{n}_1+g(A_\mu ^2A_\nu^3 -A_\nu ^2A_\mu^3 )\hat{n}_1\\
&+(\partial_\mu A_\nu ^2-\partial_\nu A_\mu^2)\hat{n}_2+g(A_\mu^3A_\nu ^1-A_\nu^3A_\mu ^1)\hat{n}_2
\end{split}
\end{equation}

\end{document}